\documentclass{PoS}

\usepackage{eurosym}
\usepackage{graphicx}
\input macros.sty

\title{Mesonic and baryonic correlation functions at fine lattice
  spacings}

\ShortTitle{Correlation functions at fine lattice spacings}

\author{Stefano Capitani, Michele Della Morte, Eric Endre{\ss},
        Andreas~J\"uttner,
	Bastian~Knippschild, \speaker{Hartmut Wittig} and Manuel
        Zambrana\\ 
        Institut f\"ur Kernphysik, University of Mainz, Becher Weg 45,
        55099 Mainz, Germany\\
        E-mail: \email{wittig@kph.uni-mainz.de}}

\abstract{
\vspace{-8cm}
We report on our on-going project to compute mesonic and baryonic two-
and three-point correlation functions in simulations using $\Nf=2$
flavours of $\rmO(a)$ improved Wilson quarks and the Wilson plaquette
action. We present performance figures for the DD-HMC algorithm on
commodity cluster hardware and discuss the issue of critical slowing
down, which is particularly pronounced for the topological charge. The
effectiveness of stochastic noise sources and Jacobi smearing are
investigated. Our preliminary results obtained at three quark masses
on $96\times48^3$ at $\beta=5.5$ imply that the lattice spacing is
about 0.06\,fm, while the smallest pion mass in the current runs is
around 360\,\MeV, which corresponds to $m_{\pi}L=5.3$.
\vspace{-13cm}{\begin{flushright} {\tt MKPH-T-09-28}\end{flushright}}
}

\FullConference{The XXVII International Symposium on Lattice Field Theory - LAT2009\\
		 July 26-31 2009\\
		 Peking University, Beijing, China}

\begin{document}

\section{Introduction}
\vspace{-0.2cm}
Despite the fact that there has been enormous recent activity in
simulating lattice QCD with dynamical quarks, the continuum limit is
still poorly understood. There are few systematic scaling studies of
hadronic quantities, and many results for phenomenologically
interesting observables have been obtained at one or two values of the
lattice spacing only. As far as control over cutoff effects is
concerned, simulations with dynamical quarks have not yet reached the
same maturity compared to the quenched approximation. As the latter is
being abandoned, one runs the risk of replacing one systematic effect
(quenching) by another. The need for having full control over all
systematics is further highlighted by the fact that lattice results
are increasingly important for providing constraints on the validity
of the Standard Model.

The work presented here is part of the CLS (``Coordinated Lattice
Simulations'') project \cite{CLSWiki}, which is aimed at generating a
set of ensembles for QCD with two dynamical flavours for a variety of
lattice spacings $(a\approx 0.04, 0.06, 0.08\,\fm)$ and volumes, such
that the continuum limit can be taken in a controlled manner.
Non-perturbatively $\rmO(a)$ improved Wilson quarks are used to
discretise the quark action. In order to tune the masses of the light
quarks towards their physical values whilst keeping the numerical
effort in the simulations at a manageable level, we employ the
deflation-accelerated DD-HMC algorithm \cite{DDHMCdefl}.

\section{Production runs}

The production runs which we carry out as part of the CLS effort are
performed on the cluster platform ``Wilson'' at the University of
Mainz, which is exclusively used for lattice
QCD\,\cite{Wilsoncluster}. It comprises 280 compute nodes, each
equipped with two AMD~2356 ``Barcelona'' processors, clocked at
2.3~GHz. Each core has one GByte of memory so that the cluster's total
memory amounts to 2.24~TBytes. Communication between nodes is realised
via an Infiniband network and switch (DDR 20+20 Gb/s full duplex). The
compute nodes are placed in water-cooled server racks. Benchmarks
based on typical QCD applications\,\cite{QCDpbm} show that the
cluster's sustained performance scales up to 3.6~TFlops, depending on
the local system size. Considering the procurement costs of
1.1~M{\euro} thus implies a cost-effectiveness of about
0.30\,\euro/MFlops (sustained). The ratio of the required cooling
capacity per compute speed amounts to 20~kW/TFlops.

We have generated configurations at $\beta=5.5$ on lattices of size
$96\cdot48^3$. Following \cite{impr:csw_nf2} we set the coefficient of
the Sheikholeslami-Wohlert term to $\csw=1.75150$. On our cluster
platform, we ran at three values of the hopping parameter
simultaneously, using 576 processor cores per job. The length of one
HMC trajectory was set to $\tau=0.5$, and the block size chosen as
$8^2\times12^2$. Further information on simulation parameters and
performance figures is listed in Table\,\ref{tab_parms}. At each value
of $\kappa$ several thousand trajectories were generated for
thermalisation. 

\begin{table}
\begin{center}
\begin{tabular}{ccccccccc}
\noalign{\vskip0.3ex}
\hline\hline\noalign{\vskip0.3ex}
Run  & Lattice & Block size & $\kappa$ & $n_0,\,n_1,\,n_2$ & time/traj. &
$P_{\rm acc}$ & $N_{\rm tr}$ & $\tau_{\rm int}[\rm plaq]$ \\ 
\noalign{\vskip0.3ex}
\hline\noalign{\vskip0.3ex}
  N3 & $96\times48^3$ & $8^2\times12^2$ & 0.13640 & 4, 5, 16 & ~763\,s
  & 0.85 & 13761 & 16(3) \\ 
  N4 &                &                 & 0.13650 & 4, 5, 20 & ~943\,s
  & 0.87 & 13104 & 14(2) \\ 
  N5 &                &                 & 0.13660 & 4, 5, 24 & 1262\,s
  & 0.86 & 12419 & 16(3) \\ 
\noalign{\vskip0.3ex}
\hline\hline
\end{tabular}
\caption{\label{tab_parms} Run parameters at $\beta=5.5$. We list the
  number of steps used in the hierarchical integration schemes
  \cite{DDHMCdefl}, the average CPU time per trajectory, acceptance
  rate $P_{\rm acc}$ and the total number of trajectories, $N_{\rm
  tr}$, generated in each run. The last column contains the integrated
  autocorrelation time of the average plaquette, obtained after
  discarding the first 3000 trajectories in each run.}
\end{center}
\end{table}

\begin{figure}
\begin{center}
\leavevmode
\includegraphics[height=5.5cm,angle=270]{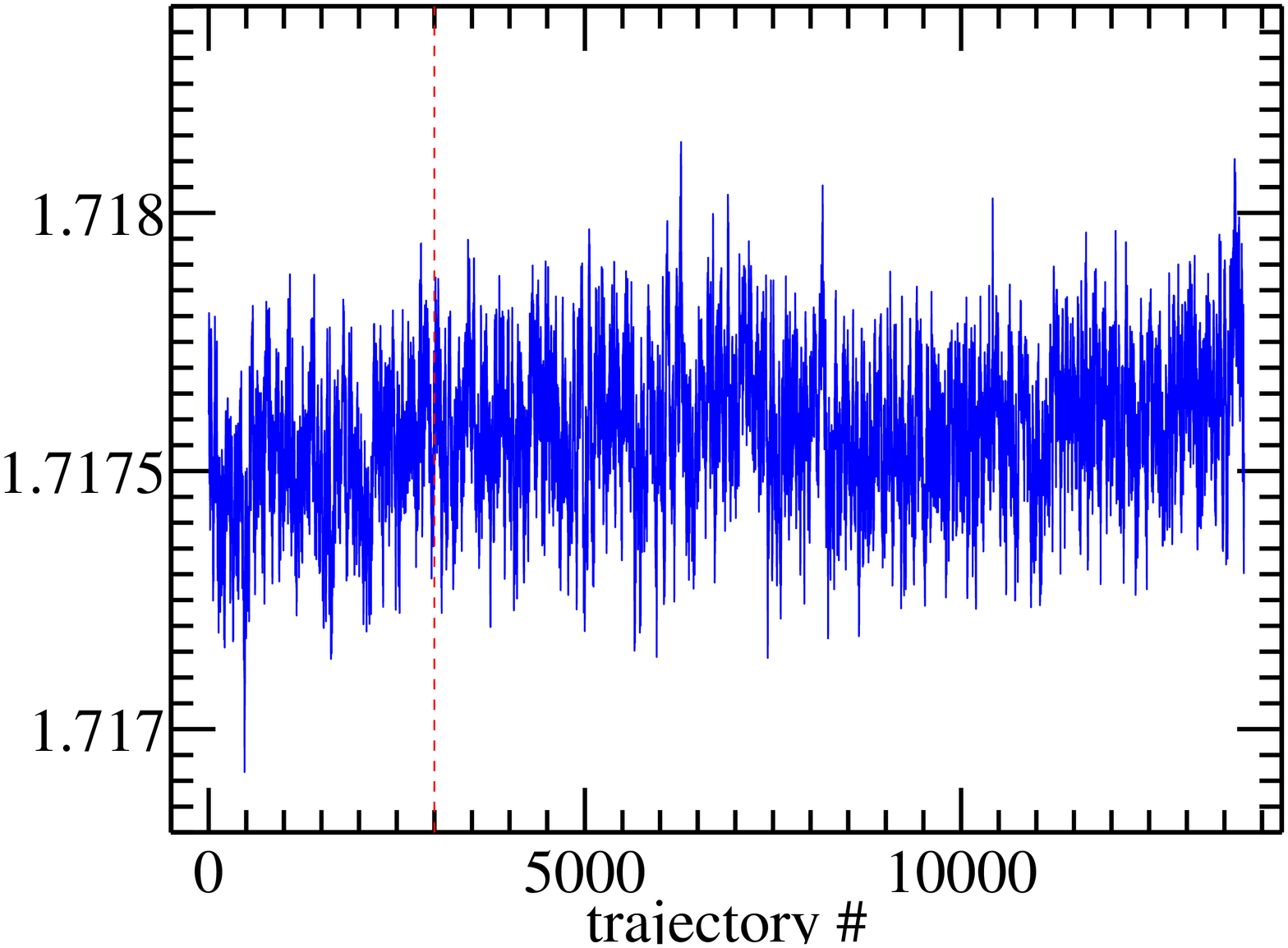}
\leavevmode\hspace{1cm}
\includegraphics[height=5.5cm,angle=270]{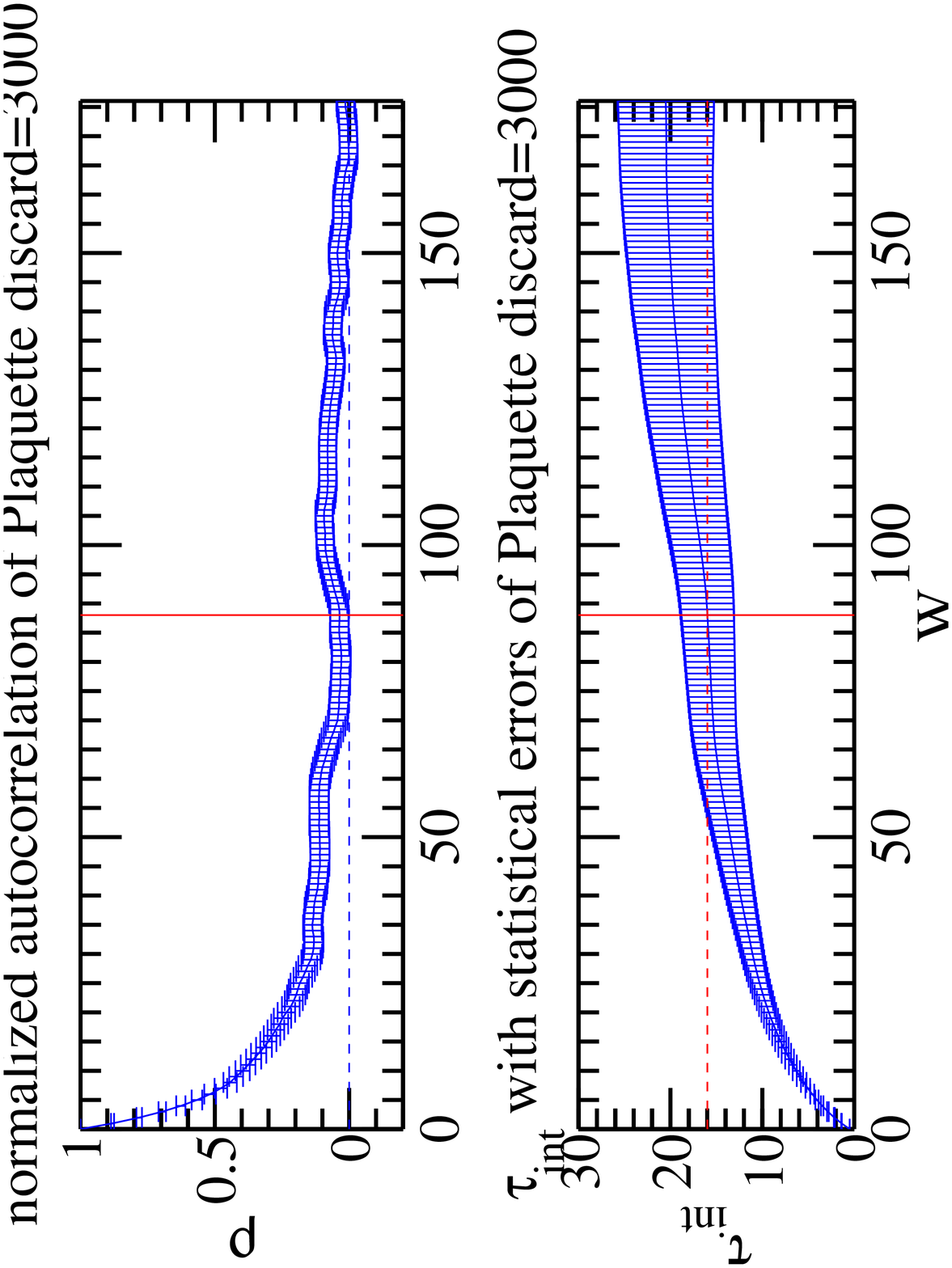}
\caption{\label{fig_MChist_plaq} {\bf Left:} Monte Carlo history
  of the average plaquette for run~N3. Trajectories to the left of the
  vertical dashed line were discarded; {\bf Right:}
  Autocorrelation functions of the average plaquette obtained after
  discarding the first 3000 trajectories.}
\vspace{0.3cm}
\includegraphics[height=5.5cm,angle=270]{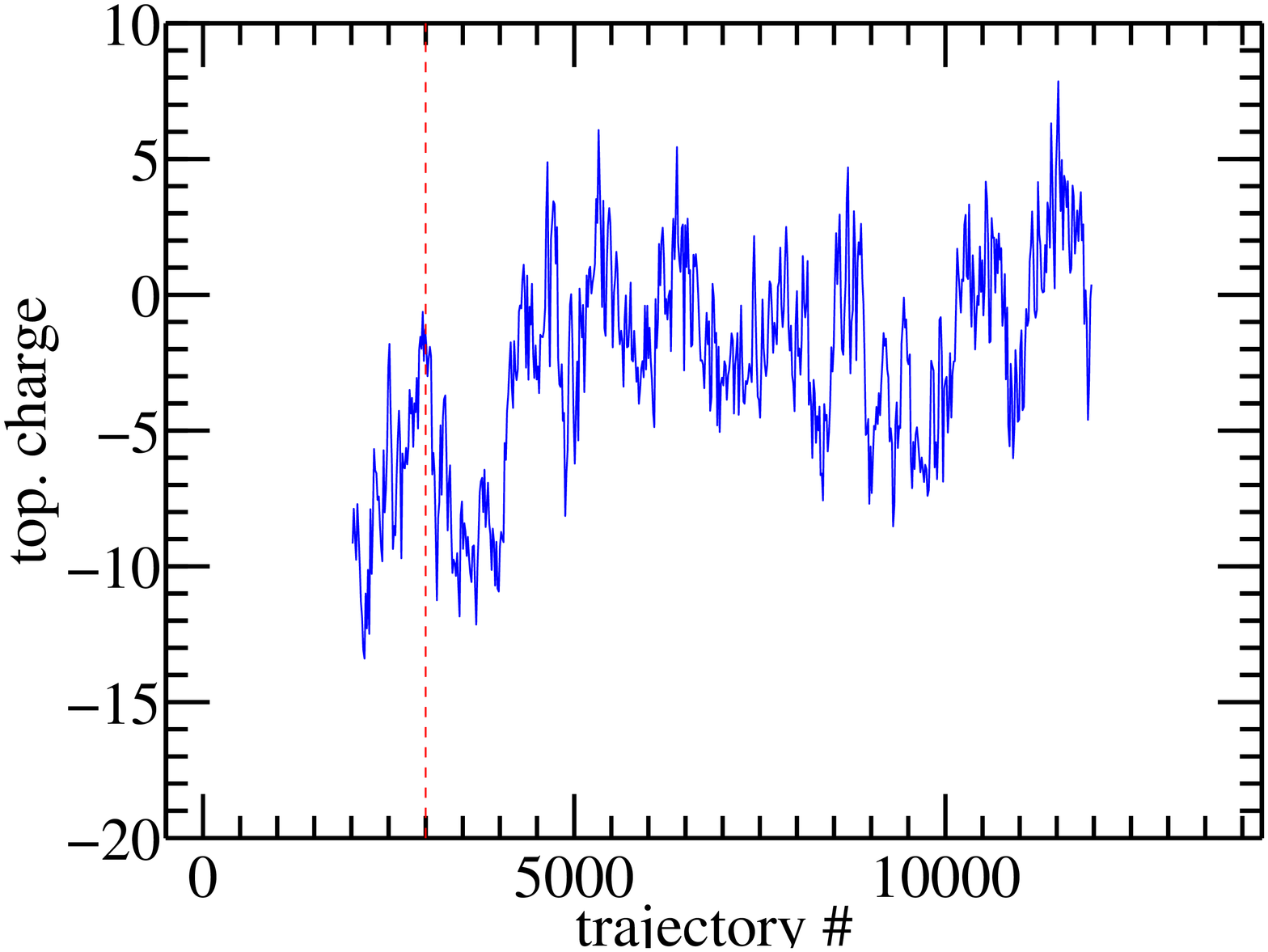}
\leavevmode\hspace{1cm}
\includegraphics[height=5.5cm,angle=270]{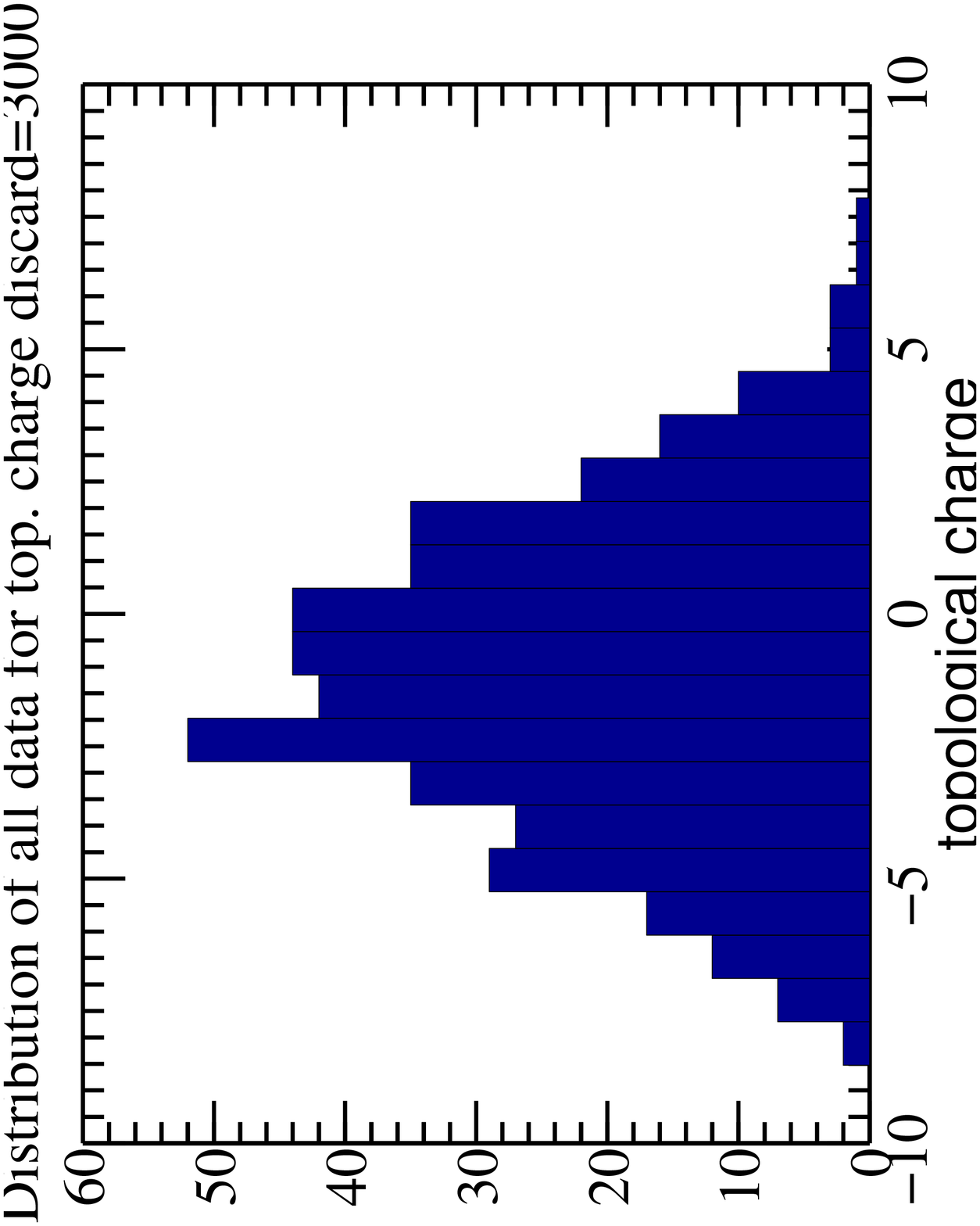}
\caption{\label{fig_MChist_top} {\bf Left:} Monte Carlo history of the
  topological charge for run~N3; {\bf Right:} Distribution of the
  topological charge after discarding the first 3000 trajectories.}
\end{center}
\end{figure}

The Monte Carlo history of the average plaquette at our largest quark
mass is shown in Fig.\,\ref{fig_MChist_plaq}. For the first 3000
trajectories a small trend in the data is observed, which is
attributed to insufficient thermalisation. Similar observations were
made at the other quark masses, and hence we discarded the first 3000
trajectories in each run. Following the method in ref.\,\cite{Wolff:2003},
the integrated autocorrelation time for the plaquette was determined,
and the resulting values are listed in Table\,\ref{tab_parms}. We
stored configurations after every 16th trajectory on disk, which
leaves us with more than 600 configurations at each quark mass.

The simulations performed as part of the CLS project revealed a severe
case of critical slowing down in the topological charge, which
manifests itself in a steep rise of the associated autocorrelation
time as a function of the lattice spacing. In particular, it was
observed\,\cite{schaefer_lat09} that at $\beta=5.7$ (which corresponds
to a lattice spacing of $a\approx0.04\,\fm$), tunnelling between
topological sectors is strongly suppressed. In
Fig.\,\ref{fig_MChist_top} we plot the Monte Carlo history for run~N3
of the topological charge, $Q=a^4\sum_{x}{\rm
tr\,}[F(x)\tilde{F}(x)]/(16\pi^2)$. With the exception of the first
3-4000 trajectories, the topological charge does fluctuate around zero
at a sizeable rate and produces a distribution which is reasonably
symmetric. Similar observations were made at the other values of the
quark masses used in our simulations. Thus, unlike the situation
encountered at the larger $\beta$-value of~5.7 \cite{schaefer_lat09}
the topological charge does not appear to be stuck in a particular
sector. While this may be accidental, we can take confidence that the
composition of our ensembles is apparently not strongly biased. We
stress that critical slowing is a general problem for lattice
simulations near the continuum, which calls for a radical treatment
like the one proposed in\,\cite{MLtriv}.

\section{Mesonic and baryonic two-point functions}

The most widely used procedure to compute quark propagators is the
source method, which amounts to solving the linear system
\be
   D\Phi=\eta,
\label{eq_source}
\ee
where $D$ is the lattice Dirac operator and $\eta$ a source vector. If
$\eta$ is chosen to be a point source, the resulting hadron
correlators can be quite noisy, with the exception of the simplest
channels such as the pion. An unambiguous identification of the
asymptotic behaviour is then quite difficult. It is not only desirable
to reduce the level of statistical noise but also to enhance the
spectral weight of the desired state in the spectral decomposition of
the correlator. In our simulations we have addressed the first problem
by comparing different stochastic noise
sources\,\cite{Eric_Dipl09}. In particular, we have implemented the
generalised ``one-end-trick''\,\cite{one_end_trick}. In order to
enhance and tune the projection properties of interpolating operators,
we have implemented several variants of Jacobi
smearing\,\cite{UKQCD_smear}.

Let $\eta$ be a random noise vector which satisfies 
\be
   \langle\langle\eta_\alpha^a(x)\eta^{b\ast}_\beta(y)\rangle\rangle =
   \delta^{(4)}(x-y)\delta^{ab}\delta_{\alpha\beta},
\ee
where double brackets denote the stochastic average. The two-point
correlation function of a quark bilinear,
$O(x)=(\psibar\Gamma\psi)(x)$ is given by
\be
   \sum_{\xvec,\yvec}\langle O(x)O(y)^\dagger\rangle =
   -\left\langle\Tr\left\{ S(x,y)\Gamma\gamma_5 S(x,y)^\dagger
   \gamma_5\tilde\Gamma \right\}\right\rangle,
\ee
where $\tilde\Gamma=\gamma_0\Gamma^\dagger\gamma_0$. The generalised
one-end-trick amounts to choosing a spin-diagonal random source
vector. More specifically, the noise source has support only on a
particular spin component $\tau$ and timeslice (e.g. $y_0=0$), viz.
\be
   \eta_\sigma^b(y) =
   \xi^b(\yvec)\delta_{0y_0}\delta_{\sigma\tau},\qquad
   \langle\langle \xi^b(\yvec)\xi^{c\ast}(\vec{z}) \rangle\rangle =
   \delta^{(3)}(\vec{y}-\vec{z})\delta^{bc}.
\ee
Solving the linear system, \eq{eq_source}, for spin component $\tau$
yields the solution vector $\Phi$, i.e.
\be
   \Phi^a_{\alpha;\tau}(x) = \sum_{{\vec{y}}}\sum_b
   \left. S^{ab}_{\alpha\tau}(x,y)\right|_{y_0=0}\xi^b(\yvec).
\ee
The correlation function is then obtained as
\be
   \sum_{\xvec,\yvec}\langle O(x)O(y)^\dagger\rangle = -
   \left\langle\sum_{\xvec} \sum_{a,\alpha,\tau} \langle\langle
   \Big[(\Gamma\gamma_5)\Phi(x)^\dagger\Big]^a_{\tau;\alpha}
   \Big[(\gamma_5\tilde\Gamma)\Phi(x)\Big]^a_{\alpha;\tau}
   \rangle\rangle\,\right\rangle,
\ee
For every ``hit'', i.e. every choice of random source one must perform
four inversions, one for each spin component $\tau$. Compared with the
point source, the numerical effort is reduced by a factor three per
hit.

\begin{figure}[t]
\begin{center}
\leavevmode
\includegraphics[height=5.5cm]{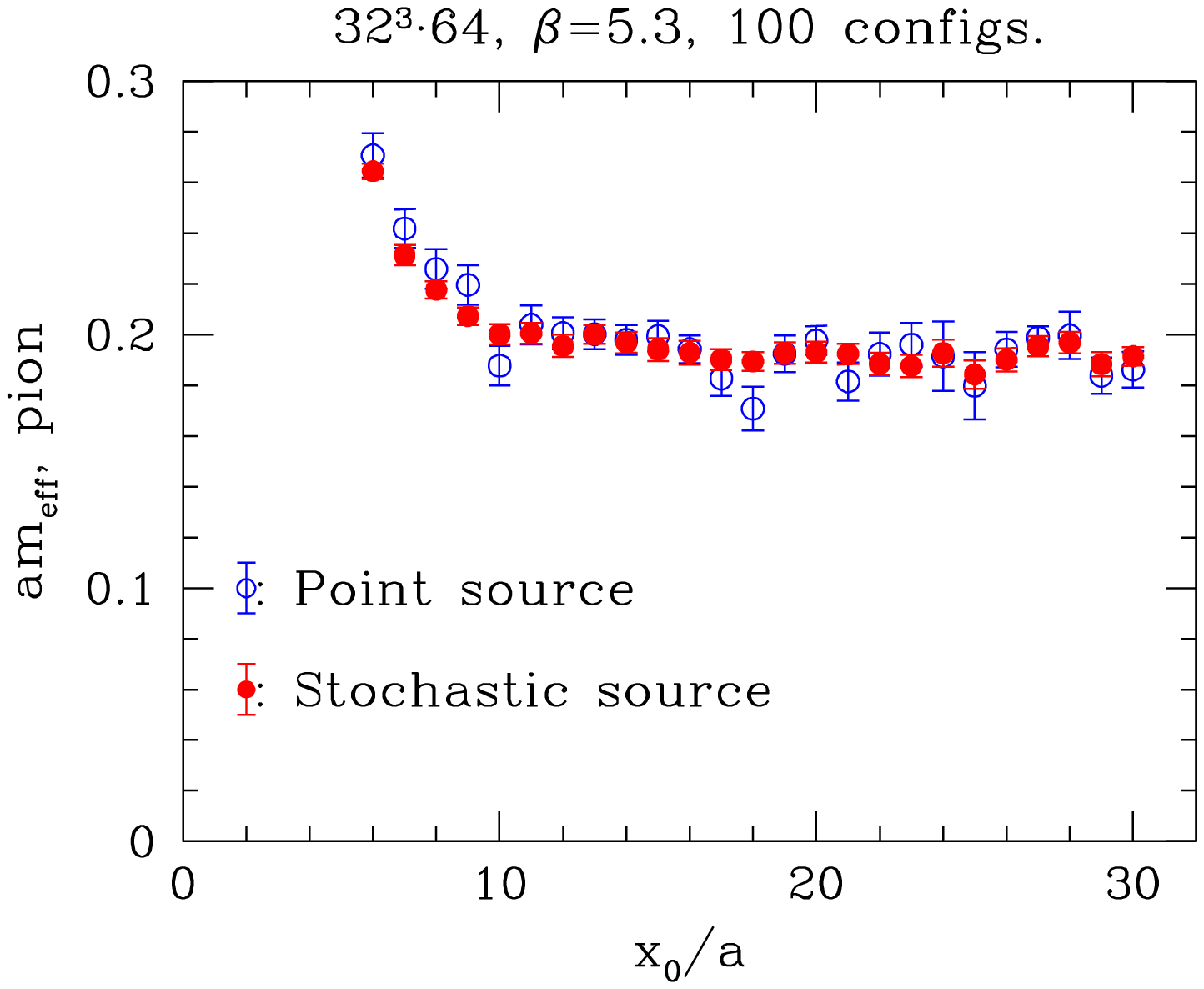}
\leavevmode\hspace{1cm}
\includegraphics[height=5.4cm]{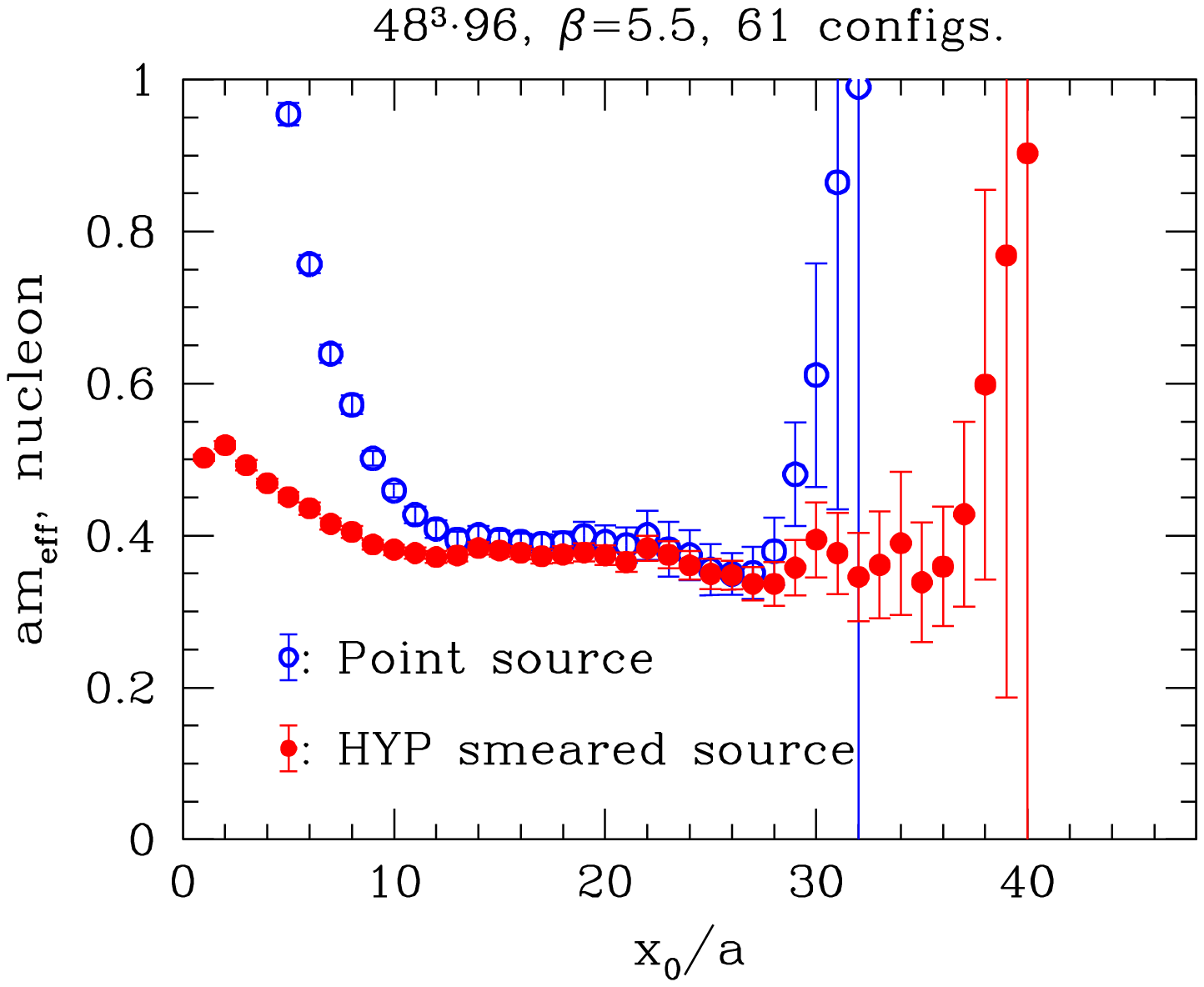}
\caption{\label{fig_effm} {\bf Left:} Effective masses in the
  pion channel obtained using a point source and spin-diagonal
  $Z_2\otimes Z_2$ noise source at fixed numerical cost; {\bf Right:}
  Effective masses for the nucleon computed on lattice~N4.} 
\end{center}
\end{figure}

In our project we have chosen $Z_2\otimes Z_2$ noise for the sources
$\xi$. In Fig.\,\ref{fig_effm} we compare the statistical signal for
the conventional point source to the generalised one-end trick for
three hits, such that the numerical cost for computing correlators for
the two source types is identical. It is seen that in the pion channel
random noise sources can lead to a significant enhancement of the
statistical signal. Further studies showed that a similar improvement
is, unfortunately, not observed in the vector channel. For baryons, we
used the method of ref.\,\cite{foley05}, but without explicit low-mode
averaging. Here, in order to reach a given statistical accuracy, the
numerical effort was at least as large as for point sources, even
after trying various dilution schemes\,\cite{Bulava_etal}, and
therefore we found the method to be practically useless for the
determination of baryonic ground state masses.

In order to enhance the projection onto the ground state in a given
channel, particularly for baryons, we have implemented Jacobi smearing
\cite{UKQCD_smear}, supplemented by ``fat'' link variables. The latter
were obtained either via the APE \cite{smear:APE} or via the
HYP\,\cite{HYP} procedure. While we found much better plateaus when
using smeared links of either type, HYP smearing appears to have a
slight advantage. In Fig.\,\ref{fig_effm} we compare effective mass
plots for the nucleon, computed using point and HYP-Jacobi smeared
sources. It is seen that not only the contribution of excited states
is reduced but that also the plateau extends to larger timeslices if
HYP-Jacobi smearing is applied, although there may be room for further
improvement via better tuning of the smearing parameters.

\section{Setting the scale}
\vspace{-0.3cm}
In order to convert the pion masses computed on our ensembles into
physical units, we must set the scale. The mass of the $\Omega$ baryon
is very well suited for this purpose, since the $\Omega$ is stable in
QCD and because it contains only strange quarks in the valence
sector. A long chiral extrapolation in the valence quark mass can thus
be avoided. For a reliable determination of the mass of the $\Omega$,
however, our simulations and analyses are not yet advanced enough. In
order to obtain preliminary values for the lattice scale, we have
therefore resorted to using the mass of the $K^\ast$-meson.

\begin{figure}
\begin{center}
\includegraphics[height=6.0cm]{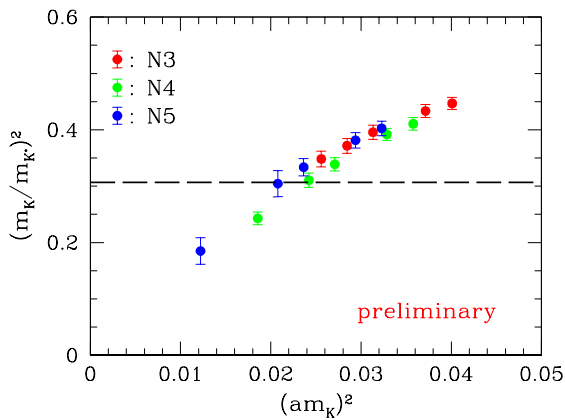}
\caption{\label{fig_interpol} The ratio $(m_K/m_{K^\ast})^2$ as a
  function of $(am_K)^2$ for the three data sets. The horizontal
  dashed lines denotes the physical ratio $m_K/m_{K^\ast}=0.554$.}
\end{center}
\end{figure}
\begin{table}
\begin{center}
\begin{tabular}{ccc}
\noalign{\vskip0.3ex}
  \hline\noalign{\vskip0.3ex}
Run & ${m_\pi}L$ & $m_\pi\,[\MeV]$ \\
\noalign{\vskip0.3ex}
  \hline\noalign{\vskip0.3ex}
N3 & 7.680(15) & 524(13) \\
N4 & 6.540(18) & 446(11) \\
N5 & 5.306(23) & 362(~9) \\
\noalign{\vskip0.3ex}
  \hline\noalign{\vskip0.3ex}
\end{tabular}
\caption{\label{tab_pionmasses} Preliminary results for pion masses in physical units.}
\vspace{-0.5cm}
\end{center}
\end{table}

To this end we have followed the procedure outlined
in\,\cite{Debbio2}: we have determined the masses of pseudoscalar and
vector mesons for degenerate and non-degenerate combinations of
quarks, where one of the masses was fixed to coincide with the sea
quark mass. We denote the masses of the generic non-degenerate
pseudoscalar and vector mesons by $m_K$ and $m_{K^\ast}$,
respectively. Their values were obtained from single-exponential fits
to the corresponding correlation functions, where the latter were
computed using stochastic sources.
The first step in the scale-setting procedure consists of
interpolating the ratio $(m_K/m_{K^\ast})^2$ as a function of
$(am_K)^2$ to the experimentally observed value of
$m_K/m_{K^\ast}=0.554$. Fig.\,\ref{fig_interpol} shows the data points
for the three ensembles. The intersection of the fit curves with the
horizontal dashed line determines the kaon mass in lattice units,
$am_K$, and thus fixes the bare mass of the strange quark at a given
value of the sea quark mass. In the second step one interpolates
$am_K$ in the sea quark mass to the reference value
$m_\pi/m_K=0.85$. Obviously, this value does not correspond to the
physical situation. However, as explained in\,\cite{Debbio2}, it
serves as a perfectly well-defined reference point, which is
sufficient for comparing data on different ensembles.
The kaon mass at the reference point is determined as $am_K|_{\rm
ref}=0.1512(38)$. After inserting the physical (isospin-averaged) kaon
mass of 495\,MeV, one obtains $a=0.0603(15)\,\fm$. This value can then
be used to convert the pion masses on the various ensembles into
physical units, which yields the values listed in
Table~\ref{tab_pionmasses}, where the combination ${m_\pi}L$ is shown
as well.

\section{Conclusions}
\vspace{-0.2cm}
Our studies have shown that large lattices at fine resolution can be
simulated efficiently on commodity clusters. In spite of a sharp
increase in the autocorrelation time of the topological charge
observed at even smaller lattice spacings\,\cite{schaefer_lat09}, the
distributions for this quantity obtained in our runs are not
pathological.
We plan to compute two- and three-point correlation functions for
mesonic and baryonic states in order to determine a variety of
observables. So far our minimum pion mass is about 360\,\MeV, and
$m_{\pi}L$ is kept larger than~5. Lowering the quark mass further, in
order to access pion masses of less than 300\,{\MeV} would necessitate
going to larger lattice sizes, if one wants to maintain the condition
$m_{\pi}L>3$.

With the currently available algorithms, i.e. while a satisfactory
solution to the problem of critical slowing down is still under
investigation, it is not worth investing more effort into the
generation of ensembles with smaller lattice spacings.
\bigskip
\par\noindent{\bf Acknowledgments} This work was supported by {\sl
Deutsche Forschungsgemeinschaft} (SFB443), {\sl Gesellschaft f\"ur
Schwerionenforschung, GSI}, and the Research Center ``Elementary
Forces and Mathematical Foundations'' funded by {\sl
  Forschungsinitiative Rheinland-Pfalz}.

\end{document}